\documentclass[12pt]{article}
\usepackage{graphicx}
\usepackage{amssymb,epsfig}
\def\bseq{\begin{subequation}}  
\def\eseq{\end{subequation}}
\def\bsea{\begin{subeqnarray}}  
\def\esea{\end{subeqnarray}}

\newcommand{\bbox}{\lower.2ex\hbox{$\Box$}}

\newcommand{\beq}{\begin{equation}}
\newcommand{\eeq}{\end{equation}}
\newcommand{\bea}{\begin{eqnarray}}
\newcommand{\eea}{\end{eqnarray}}
\newcommand{\ena}{\end{eqnarray}}

\newcommand {\non}{\nonumber}

\renewcommand{\a}{\alpha}
\renewcommand{\b}{\beta}

\newcommand{\g}{\gamma}

\newcommand{\D}{\Delta}

\newcommand{\m}{\mu}

\begin{document}


\vskip 0.75 cm
\renewcommand{\thefootnote}{\fnsymbol{footnote}}
\centerline{\Large \bf An $\mathcal{N}=1$ Superfield Action for M2 branes }
\vskip 1.5 cm

\centerline{{\bf Andrea Mauri${}^{1}$\footnote{Andrea.Mauri@mi.infn.it} and Anastasios C. Petkou${}^{1}$\footnote{petkou@physics.uoc.gr} 
}}
\vskip .5cm
\centerline{${}^1$\it Department of Physics,
University of Crete, 
Heraklion 71003, Greece}

\vskip 1cm

\setcounter{footnote}{0}
\renewcommand{\thefootnote}{\arabic{footnote}}

\begin{abstract}
We present an octonionic ${\cal N}=1$ superfield action that reproduces in components the action of Bagger and Lambert for M2 branes. By giving an expectation value to one of the scalars we obtain the maximally supersymmetric superfield action  for  D2 branes.
\end{abstract}

\section{Introduction}

Despite its importance for $M$-theory and higher-spin gauge theory
 holography\footnote{For a general discussion see \cite{Tassos}.}
 AdS$_4$/CFT$_3$ correspondence is essentially unexplored in
 comparison to its AdS$_5$/CFT$_4$ counterpart. Presumably, the main
 reason has been the lack so far of a manifestly maximally
 superconformal invariant 2+1 dimensional theory similar in status to
 ${\cal N}=4$ SYM in 3+1 dimensions. It seems possible that this
 obstacle has been overcome by the proposal of Bagger and Lambert
 \cite{BL1,BL2,BL3,Gustav} of a ${\cal N}=8$ superconformal theory in
 2+1 dimensions, with a non-standard gauge structure based on
 3-algebras. This theory has attracted very much interest in the past
 few months and many interesting results have appeared.  Multiple M2
 branes have been discussed in \cite{Papag1,Berman,Raams,Morozov,
 Distler, Gran,Roo,Song,Bandres,Gomis2,Gomtwo,Macca}. The relation of M2 to D2
 branes was elucidated in
 \cite{Papag1,Tong,Ho2,Benvenuti,Ho3,Papag2}. Algebraic aspects of
 3-algebras have been discussed in
 \cite{Gustavsson3,Ho,Papad1,Papad2,Gomis,Lin,Passerini,Figueroa1,Figueroa2}.
 General aspects of three-dimensional Chern-Simons theories and
 extensions have been discussed in \cite{Schwarz,Witten} and also in
 \cite{ Park1,Park2,Hosomichi,Shimada,Morozov2,Banerjee,Gustavsson,Park3}.
 Very recently, an interesting class of $U(N)\times U(N)$ Chern-Simons
 theories that may describe multiple M2 branes have been discussed in
 \cite{Aharony,Gomis2,Benna}.

In this short note, we present an ${\cal N}=1$ superfield action whose component expansion gives the BL theory for a 3-algebra with totally antisymmetric structure constants $f^{abcd}$. We use real three-dimensional superfields both for the matter as well as for the Chern-Simons part of the action. The crucial point is the use of the octonionic self-dual tensor in the construction of the real superpotential. In this way, the superpotential is only manifestly $SO(7)$ invariant. However, for specially chosen couplings, the component action coincides with the BL action, and hence full $SO(8)$ symmetry is restored. We believe that octonions will play a fundamental role in future studies of AdS$_4$/CFT$_3$. 

Our motivation comes in part from corresponding studies in AdS$_5$/CFT$_4$ where the ${\cal N}=1$ formulation of ${\cal N}=4$ SYM has been an extremely efficient tool for studies of anomalous dimensions, non-renormalization properties and integrability. We believe that our ${\cal N}=1$ action will be similarly useful in this case too. 


As a simple test for our action we follow \cite{Papag1} and demonstrate that giving an expectation value in one of the scalar supefields, our action yields the maximally supersymmetric YM theory in 2+1 dimensions, as it should.

\section{The ${\cal N}=1$ superfield action}

We consider eight real ${\cal N}=1$ superfields\footnote{Earlier works on the superfield formulation of Chern-Simons theories coupled to matter include \cite{Avdeev1,Avdeev2}.} as \beq \Phi^{I}_a =
\phi^I_a+\theta^{\a,\dot{8}}
\hat{\Gamma}^{I}_{\dot{8}A}\psi^A_{\a,a}-\theta^2F^I_a\,,\,\,\,
I,A=1,2,..,8.  \eeq where $a$ denotes the index of the three-algebra
algebra with structure constants $f^{abcd}$. We use the $SO(8)$
triality tensor $\Gamma^I_{A\dot{A}}$, $I,\dot{A},A=1,2,..,8$. In the
representation where $\hat{\Gamma}^I_{\dot{8}A}=-\delta^I_A$ (see appendix
B), choosing the superspace coordinate to point in the $\dot{8}$
direction we have essentially made equivalent the vector and one of
the two spinorial representations of $SO(8)$. We use the notation of
\cite{superspace} summarized along with other useful
relations and conventions in the Appendix A.

Our superfield action couples the matter superfields to a Chern-Simons gauge superfield $\Gamma^\a_{ab}$ in the Wess-Zumino gauge, that carries two gauge group indices $a,b$  
\begin{eqnarray}
S & = & \int d^3 x \, d^2 \theta \,\Big[\, 2\gamma \left(
D^{\alpha}\Phi^{I}_{d} - \, f
^{abc}_{\phantom{abc}d}\,\Gamma^{\alpha}_{\phantom{\alpha}ab}
\Phi^{I}_{c} \right)^2 +\, \,\alpha \,f^{abcd}\,(D^{\alpha}
\Gamma^{\beta}_{\phantom{\beta}ab})(D_{\beta} \Gamma_{\alpha \,cd}) +
\,\non\\
\label{action}
 & & + \beta \,f^{cda}_{\phantom{cda}g} f^{efgb}\, (D^{\alpha}
\Gamma^{\beta}_{\phantom{\beta}ab})\Gamma_{\alpha \,c d}\Gamma_{\beta
\, e f} \,+\, k \,f^{abcd} \,C_{IJKL}
\,\Phi^{I}_{a}\Phi^{J}_{b}\Phi^{K}_{c}\Phi^{L}_{d}
\,\Big] \,.
\end{eqnarray}
Apart from the overall normalization, the only adjustable parameter is the coupling constant of the real superpotential. Nevertheless, we keep all  coefficients $\a,\b,\g,k$ arbitrary  having in mind possible generalizations of the action (\ref{action}). 

The crucial point is the use of the  self-dual\footnote{Similar result can be obtained using the anti self-dual tensor.} eight dimensional tensor $C_{IJKL}$, $I,J,K,L=1,2,..8$ that describes the embeddings of $SO(7)$ into $SO(8)$ \cite{Nicolai,Gunaydin,Floratos,Sfetsos}. Its properties are briefly recalled in the Appendix B. Hence, the presence of the superpotential implies that (\ref{action}) has only $SO(7)$ manifest global symmetry.  Our strategy is to fix the coefficients in (\ref{action}) by comparing the resulting component action with the one of BL. In this way, the $SO(7)$ symmetry is enhanced to  $SO(8)$ and hence we achieve maximal supersymmetry. 

We detail next the various projections.

\noindent A) {\it Scalar kinetic terms}
\bea
&& \left.D^2 \left[2\gamma \left( D^{\alpha}\Phi^{I}_{d} - \, f
^{abc}_{\phantom{abc}d}\,\Gamma^{\alpha}_{\phantom{\alpha}ab}
\Phi^{I}_{c} \right)^2 \right]\right|_{\theta=0} = \vspace{0,3cm}\non\\
&& = -2\gamma\, F^{I}_{\,d} F^{I \, d} -2\gamma\,
\phi^{I}_{\,\,d} \,\Box \,\phi ^{I \, d} - 2 \gamma\,i
\,\psi^{\alpha\,I}_{\,\,d} \partial^{\,\,\beta}_{\alpha}
\psi_{\beta}^{\,\,I\,d} + 2 \gamma  f^{abc}_{\phantom{abc}d}
f^{efgd}\, A^{\mu}_{\,\,ab}
A_{ \mu\,ef}\, \phi^{I}_{\,\,c} \phi^{I}_{\,\,g} -
\non \\
&&\,\,\,\,\, - \gamma \,
f^{abc}_{\phantom{abc}d} \left[-\phi^{I}_{\,\,c}
\left(\gamma_{\mu} \right)^{\alpha}_{\,\,\gamma}
A^{\mu}_{\,\,ab}\partial^{\gamma}_{\,\,\alpha}\,\phi^{I\, d} - \left(
\partial^{\alpha}_{\,\,\gamma} \phi^{I\,d}\right)
\left(\gamma_{\mu} \right)^{\gamma}_{\,\,\alpha}
A^{\mu}_{\,\,ab}\,\phi^{I\,}_{\, c}\right. +   \non \\ \vspace{1cm}
&& \hspace{2.5cm}+ 2 i \,\psi^{\gamma \,I}_{\,\,c}
\left(\gamma_{\mu}\right)_{\gamma}^{\,\,\alpha} A^{\mu}_{\,\,ab} \psi^{I
\, d}_{\alpha}+4
 \lambda^{\alpha}_{\,\,ab}\, \psi^{\,I
\, d}_{\alpha}\phi^I_c \, \Big] = \non\\ 
\label{ScalKin}&&= 2 \gamma
\Big( \nabla^{\mu} \phi^{I}_{\,\,d} \Big) \Big(
\nabla_{\mu} \phi^{I \,d} \Big) - 2 i \,\gamma \, \psi^{
\alpha\,I}_{\,\,d} \nabla^{\,\,\beta}_{\alpha}\,\psi^{I\,d}_{\beta} -
2 \gamma\, F^{I}_{\,\,d} F^{I \, d} - 4
\gamma f^{abc}_{\phantom{abc}d} \lambda^{\alpha}_{\,\,ab}\, \psi^{\,I
\, d}_{\alpha}\phi^I_c  \, ,
\eea
where 
\beq
 \nabla_{\mu} \phi^{I}_{d} \equiv \partial_{\mu} \phi^{I \, d} -
\,f^{abc}_{\phantom{abc}d}\, A_{\mu\,ab} \,\phi^{I\, c}\,.
\eeq

\noindent
B) {\it Chern-Simons terms}:
\bea
&&\left. D^2\left[ \,\alpha \,f^{abcd}\,(D^{\alpha}
\Gamma^{\beta}_{\,\,ab})(D_{\beta} \Gamma_{\alpha \,cd}) +\,
\beta\,f^{cda}_{\phantom{cda}g} f^{efgb}\, (D^{\alpha}
\Gamma^{\beta}_{\,\,ab})\Gamma_{\alpha \,c d}\Gamma_{\beta \, e
f}\right] \right|_{\theta=0} =
\non \\ 
&&\label{CSterms}
\hspace{-.3cm}= 4\,\alpha\, f^{abcd} \lambda^{\alpha}_{\,\,ab}
\lambda_{\alpha\,cd} - 4  \alpha\,f^{abcd} \,\epsilon^{\mu \nu
\rho}\,A_{\mu\,ab}\,\partial_{\nu}A_{\rho\,cd} - 2
\beta\, f^{cda}_{\phantom{cda}g}f^{efgb}\,\epsilon^{\mu
\nu \rho}\,A_{\mu\,ab}\,A_{\nu\,cd}\,A_{\rho\,ef}\,. 
\eea

\noindent C) {\it Superpotential}:
\beq\label{Spotential} 
\left. D^2 \Big[ k \,f^{abcd} \, C_{IJKL}
\,\Phi^{I}_{\,\,a}\Phi^{J}_{\,\,b}\Phi^{K}_{\,\,c}
\Phi^{L}_{\,\,d}\Big]\right|_{\theta = 0} = k\,f^{abcd}\, C_{IJKL}
\Big[6\, \psi^{\alpha I}_{\,\,a} \psi_{\alpha \,b}^{J}
\phi^{K}_{\,\,c} \phi^{L}_{\,\,d} + 4\,F^I_{\,\,a}\phi^{J}_{\,\,b}
\phi^{K}_{\,\,c} \phi^L_{\,\,d} \Big]\,.
\eeq

\noindent The equations of motion for the auxiliary fields read
\begin{eqnarray}
\label{Feom}
F^{I a} & = & \frac{k}{\gamma}\, \,f^{abcd}\,C^I_{\phantom{I}JKL} \,
\phi^{J}_{\,\,b} \phi^{K}_{\,\,c} \phi^{L}_{\,\,d} \,,\\
\label{leom}
\lambda_{\alpha\,\,ab} & = & \frac{\gamma}{2 \alpha}\phi^I_{\, a}
\psi^I_{\alpha \, b} \,.
\end{eqnarray}\\
Using  (\ref{Feom}) and (\ref{leom}), the terms contributing to the potential from A), B) and C) that involve the auxiliary fields  give
\bea
&& \hspace{-1cm}-\frac{\gamma^2}{\alpha} f^{abcd} \psi^{\alpha\, I}_{ a}\phi^I_{\,
b}\psi^J_{\alpha \, c}\phi^J_{\,d }+ \frac{2 k^2 }{\gamma} \,
f^{bcda}\, f^{efg}_{\phantom{efg}a} \,C^I_{\phantom{I}OMN}\, C_{IJKL}
\,\phi^{O}_{\,\,e} \phi^{M}_{\,\,f} \phi^{N}_{\,\,g}
\phi^{J}_{\,b}\phi^{K}_{\,c}\phi^{L}_{\,d} = \nonumber \\
&& - \frac{
\gamma^2}{\alpha} f^{abcd} \psi^{\alpha\, I}_{ a}\phi^I_{\,
b}\psi^J_{\alpha \, c}\phi^J_{\,d } + \frac{12 k^2 }{\gamma} \,
f^{bcda}\, f^{efg}_{\phantom{efg}a} \,\phi^{J}_{\,\,e}
\phi^{K}_{\,\,f} \phi^{L}_{\,\,g}
\phi^{J}_{\,b}\phi^{K}_{\,c}\phi^{L}_{\,d}+\, \nonumber \\
&&\hspace{2cm} -\, \frac{18 k^2
}{\gamma} \, f^{bcda}\, f^{efg}_{\phantom{efg}a} \,
C_{KLOM} \,\phi^{J}_{\,\,e} \phi^K_{f} \phi^L_{g}
\phi^{O}_{\,b}\phi^{M}_{\,c}\phi^{J}_{\,d} =\nonumber \\
\label{auxpot}
=&& -\frac{ \gamma^2}{\alpha} f^{abcd} \psi^{\alpha\, I}_{
a}\phi^I_{\, b}\psi^J_{\alpha \,
c}\phi^J_{\, d} + \frac{12 k^2}{\gamma}
 \,\textrm{Tr}\Big([\phi^J,\phi^K,\phi^L],[\phi^J,\phi^K,\phi^L]\Big)\,.
\eea
To arrive at the last line we have used the contraction of the self-dual form \cite{Nicolai,Gunaydin,Sfetsos}
\beq
C^I_{\phantom{I}OMN}\,C^{IJKL} = \delta^{[J}_{\,O}
\delta^{\,K}_{\,M} \delta^{L]}_{\,N} - 9
\,C^{KL}_{\phantom{KL}OM}\delta_{N}^{J}\,.
\eeq
It is also crucial that the third line in (\ref{auxpot}) vanishes. This can be shown for a general 3-algebra with totally antisymmetric structure constants $f^{abcd}$. Indeed, using the fundamental identity 
\beq
\label{fundid}
f^{bcda}\,f^{efg}_{\phantom{efg}a}=f^{efda}\,f^{bcg}_{\phantom{bcg}a}+f^{efba}\,f^{cdg}_{\phantom{cdg}a}+f^{efca}\,f^{dbg}_{\phantom{dbg}a}
\eeq
in the relevant part of the  third line of (\ref{auxpot}) we obtain after some relabeling 
\bea
 f^{bcda}\, f^{efg}_{\phantom{efg}a} \,
C_{KLOM} \,\phi^{J}_{\,\,e} \phi^K_{f} \phi^L_{g}
\phi^{O}_{\,b}\phi^{M}_{\,c}\phi^{J}_{\,d}&=& -2f^{bcga}\,f^{efd}_{\phantom{efd}a}C_{KLOM} \phi^{J}_e\phi^K_f\phi^L_g\phi^O_b\phi^M_d\phi^J_c \nonumber \\
 &=& -2f^{bcda}\,f^{efg}_{\phantom{efg}a}C_{KLOM} \phi^{J}_e\phi^K_f\phi^L_g\phi^O_b\phi^M_c\phi^J_d\,,
 \eea
where we have also used the antisymmetry of $C_{KLOM}$. Hence this term vanishes.

\noindent Putting everything together we arrive at the component action
\begin{eqnarray} 
\label{action1} S& = & \int d^3 x \Big[2 \gamma\Big( \nabla^{\mu}
 \phi^{I}_{\,\,d} \Big) \Big( \nabla_{\mu} \phi^{I \,d} \Big) - 2
\gamma i \psi^{ \alpha\,I}_{\,\,d}
\nabla^{\,\,\beta}_{\alpha}\,\psi^{I\,d}_{\beta} +\non \\ 
& & -
4\,\alpha\,f^{abcd} \,\epsilon^{\mu \nu
\rho}\,A_{\mu\,ab}\,\partial_{\nu}A_{\rho\,cd} - 2  \beta \,
f^{cda}_{\phantom{cda}g}f^{efgb}\,\epsilon^{\mu \nu
\rho}\,A_{\mu\,ab}\,A_{\nu\,cd}\,A_{\rho\,ef} +  \\&& \hspace{-1cm}+ 6
\,k\,f^{abcd}\, (C_{IJKL}+\frac{\gamma^2}{6 k \alpha}\delta_{IK}
\delta_{JL})\, \psi^{\alpha I}_{\,\,a} \psi_{\alpha \,b}^{J}
\phi^{K}_{\,\,c} \phi^{L}_{\,\,d} + \frac{12 k^2}{\gamma} \non
\textrm{Tr}\Big([\phi^I,\phi^J,\phi^K],[\phi^I,\phi^J,\phi^K]\Big)
\Big]\,. \non
\end{eqnarray}
Now, in our representation we have
\begin{equation}
\qquad \psi^{\alpha\,I}_a = \hat{\Gamma}^{I}_{\dot{8}A}\psi^{\alpha \, A}_{a}=
-\delta^{I}_{A}\psi^{\alpha \, A}_{a}  \,, 
\end{equation}
so that for the spinor kinetic term we have
\beq
 - 2 \gamma i \psi^{ \alpha\,I}_{\,\,d}
\nabla^{\,\,\beta}_{\alpha}\,\psi^{I\,d}_{\beta} = - 2 \gamma i \psi^{
\alpha\,A}_{\,\,d} \nabla^{\,\,\beta}_{\alpha}\,\psi^{A\,d}_{\beta}\,.
\eeq
Then,  choosing
\beq \label{vallon}
\alpha= -\frac{1}{8}\qquad \beta= -\frac{1}{6}
\qquad \gamma=-\frac{1}{4} \qquad k = -\frac{1}{24}\,,
\eeq
we obtain the action
\begin{eqnarray} \label{BL}
S & = & \int d^3 x \Big[-\frac{1}{2}\Big( \nabla^{\mu}
 \phi^{I}_{\,\,d} \Big) \Big( \nabla_{\mu} \phi^{I \,d} \Big) +
 \frac{i}{2} \psi^{ \alpha\,A}_{\,\,d}
 \nabla^{\,\,\beta}_{\alpha}\,\psi^{A\,d}_{\beta} + \non \\ & & +
 \frac{1}{2} f^{abcd} \,\epsilon^{\mu \nu
 \rho}\,A_{\mu\,ab}\,\partial_{\nu}A_{\rho\,cd} + \frac{1}{3}
 f^{cda}_{\phantom{cda}g}f^{efgb}\,\epsilon^{\mu \nu
 \rho}\,A_{\mu\,ab}\,A_{\nu\,cd}\,A_{\rho\,ef} +  \\&& \hspace{-1cm}-
 \frac{1}{4}\,f^{abcd}\, (C_{IJKL} + \delta_{IK}
 \delta_{JL} - \delta_{IL}\delta_{JK})\, \psi^{\alpha I}_{\,\,a} \psi_{\alpha \,b}^{J}
 \phi^{K}_{\,\,c} \phi^{L}_{\,\,d} -  \frac{1}{12}
 \textrm{Tr}\Big([\phi^I,\phi^J,\phi^K],[\phi^I,\phi^J,\phi^K]\Big) \,.\non
\end{eqnarray}
This coincides exactly with  the Bagger and Lambert action given in \cite{BL2}.  To see that, notice that  in our
  notations we use the purely imaginary charge conjugation matrix
  $C$ to raise and lower spinor indices. Therefore for a real Majorana
  spinor $\psi$ we have the identifications
\beq
(\bar{\psi})^\alpha = (\psi^T
C)^\alpha = C^{\alpha \beta}\psi_\beta = \psi^\alpha =
i(\bar{\psi}_{BL})^\alpha\,,
\eeq
 and also to match the fermion kinetic term 
 \beq
 i
\,\psi^{\alpha} (\gamma^{\mu}\partial_\mu)_{\alpha}^{\,\,\beta}\,
\psi_{\beta} = \left(\bar{\psi}_{BL}\right)^{\alpha} (\gamma^{\mu}\partial_\mu)_{\alpha}^{\,\,\,\b}\, \left(\psi_{BL}\right)_{\beta} = i\,\left(\bar{\psi}_{BL}\right)^{\alpha} (\gamma^{\mu}_{BL}\partial_\mu)_{\alpha}^{\,\,\,\b}\, \left(\psi_{BL}\right)_{\beta}  \,,\eeq
that requires (note the position of the $\gamma$-matrix indices)
\beq
\left(\gamma^\m_{BL}\right)_\a^{\,\,\b}=i\,\left(\gamma^\m\right)_\a^{\,\,\b}\,.
\eeq

\noindent Our final superspace action is then written as
\begin{eqnarray}
S & = & \int d^3 x \, d^2 \theta \,\Big[\, -\frac{1}{2} \left(
D^{\alpha}\Phi^{I}_{d} - \, f
^{abc}_{\phantom{abc}d}\,\Gamma^{\alpha}_{\phantom{\alpha}ab}
\Phi^{I}_{c} \right)^2 - \frac{1}{8} \,f^{abcd}\,(D^{\alpha}
\Gamma^{\beta}_{\phantom{\beta}ab})(D_{\beta} \Gamma_{\alpha \,cd}) +
\,\non\\
\label{action_final}
 & & -\frac{1}{6} \,f^{cda}_{\phantom{cda}g} f^{efgb}\, (D^{\alpha}
\Gamma^{\beta}_{\phantom{\beta}ab})\Gamma_{\alpha \,c d}\Gamma_{\beta
\, e f} \,-\frac{1}{24} \,f^{abcd} \,C_{IJKL}
\,\Phi^{I}_{a}\Phi^{J}_{b}\Phi^{K}_{c}\Phi^{L}_{d}
\,\Big] \,.
\end{eqnarray}

\section{M2 to D2}
As a simple test for our action (\ref{action_final}) we give an
expectation value to one of the scalars. Following
\cite{Papag1} we expect that the resulting action will be the
maximally supersymmetric YM theory in 2+1 dimensions for the gauge group $SU(2)$. We split
the $I=1,..,8$ index of the scalar superfield as  $I\mapsto (i,8)$ with
$i=1,...,7$ and the $SO(4)$ index $\hat{a}=1,...,4$ as  $\hat{a}\mapsto (a,x)$ with
$a=1,...,3$. Then we give an expectation value to the scalar
superfield which we identify with the dimensionful coupling constant of the 2+1 SYM as
\beq
\langle \Phi^8_x \rangle = g_{YM}\,.
\eeq
For the spinor superfield we define
\begin{eqnarray}
\Gamma^{\alpha}_{ax} & \equiv & A^{\alpha}_a \,, \\
\epsilon^{abc} \Gamma^{\alpha}_{ab} & \equiv & B^{\alpha}_c \,.
\end{eqnarray}
We rewrite the superspace action in terms of the new fields and indices. \\\\
\noindent
i) {\it Kinetic terms}: \bea && 2\gamma(D^{\alpha}\Phi^I_{\hat{d}} -
\epsilon^{\hat{a}\hat{b}\hat{c}}_{\phantom{abc}\hat{d}}
\Gamma^{\alpha}_{\,\,\hat{a}\hat{b}}\Phi^I_{\hat{c}})^2 = \gamma
\left[ \nabla^{\alpha}\Phi^i_d \nabla_{\alpha}\Phi^{id} +
\nabla^{\alpha}\Phi^8_d \nabla_{\alpha}\Phi^{8d} +
D^{\alpha}\Phi^i_{x} D_{\alpha}\Phi^{ix} + \right.\non\\
&&\hspace{-1cm}\left. + B_d^{\alpha}(2
\Phi^i_{x}\nabla_{\alpha}\Phi^{id} + 2 g_{YM}
\nabla_{\alpha}\Phi^{8d}-2\Phi^{id}D_{\alpha}\Phi^{ix} ) +
B^{\alpha}_d B^d_{\alpha}(g^2_{YM} + \Phi^i_{x}\Phi^i_{x}) + B^{\alpha
c}B_{\alpha}^g \Phi^i_{c}\Phi^i_g \right]\,.  \eea where we defined
the gauge covariant derivative \bea \label{covdev} &&
\nabla^{\alpha}\Phi^i_d = D^\alpha \Phi^i_d - 2 \epsilon^{bcd}
A^\alpha_b\phi^i_c \eea ii) {\it Superpotential term}: \beq k
\,\epsilon^{\hat{a}\hat{b}\hat{c}\hat{d}}\,C_{IJKL}
\,\Phi^{I}_{\hat{a}} \Phi^{J}_{\hat{b}} \Phi^{K}_{\hat{c}}
\Phi^{L}_{\hat{d}} = 4\,k\, \epsilon^{bcd}\,c_{jkl}\,
g_{YM}\,\Phi^{j}_{b} \Phi^{k}_{c} \Phi^{l}_{d}+\cdots\,, \eeq where
the dots indicate subleading terms in the large $g_{YM}$ limit which
we have discarded.\\

\noindent iii) {\it Chern-Simons terms}:
\bea
&&\alpha \,\epsilon^{\hat{a}\hat{b}\hat{c}\hat{d}}\,
(D^{\alpha}\Gamma^{\beta}_{\,\,\hat{a}\hat{b}})
D_{\beta}\Gamma_{\alpha\,\,\hat{c} \hat{d}} + \beta
\,\epsilon^{\hat{c}\hat{d}\hat{a}}_{\phantom{cda}\hat{g}}\,
\epsilon^{\hat{e}\hat{f}\hat{g}\hat{b}}\,(D^{\alpha}
\Gamma^{\beta}_{\,\,\hat{a}\hat{b}})\Gamma_{\alpha\,\,\hat{c}\hat{d}}
\Gamma_{\beta\,\,\hat{e}\hat{f}}= 4 \alpha \,B^{\beta
d}(D^{\alpha}D_{\beta} A_{\alpha d}) +\non\\
&&
+ 2 \beta\, \epsilon^{dag}\,B^{\beta}_g \,\Big[
2(D^{\alpha}A_{\beta a})A_{\alpha\,d} + (D_{\beta}A^{\alpha}_a)A_{\alpha
d} + (D^{\alpha}A_{\alpha a})A_{\beta d}\Big] + \frac{\beta}{2}\,
\epsilon^{abc}(D^{\alpha}B_c^{\beta}) B_{\alpha a} B_{\beta b}=\non \\
&&= B^{\beta}_d\Big[4 \alpha \, D^{\alpha}D_{\beta} A_{\alpha}^d + 6
\beta{} \epsilon^{dba}(D^{\alpha}A_{\beta a})A_{\alpha\,b} \Big] +
\frac{\beta}{2}\, \epsilon^{abc}(D^{\alpha}B_c^{\beta}) B_{\alpha a}
B_{\beta b}\,.
\eea
To arrive in the the last line we have used  the fact that in the Wess-Zumino
gauge $D^\alpha \Gamma_{\alpha}$ is vanishing and
$D_\alpha\Gamma_{\beta}$ can be symmetrized. 

Next, we derive the
equations of motions for the $B$ auxiliary superfield neglecting 
the terms cubic in $B$ and also terms of the form $B^2\Phi^2$, as in \cite{Papag1}. We
thus obtain
\beq
B_{\alpha\,d} =-\frac{1}{2 \gamma g^2_{YM}}\Big[ W_{\alpha\,d}  + 2\gamma (
\Phi^i_{x}\nabla_{\alpha}\Phi^{i}_d +  g_{YM}
\nabla_{\alpha}\Phi^{8}_d-\Phi^{i}_dD_{\alpha}\Phi^{ix})\Big] \,,
\eeq
where we defined
\beq
W_{\beta}^d =4 \alpha \, D^{\alpha}D_{\beta} A_{\alpha}^d + 6
\beta{} \epsilon^{dba}(D^{\alpha}A_{\beta a})A_{\alpha\,b}\,.
\eeq
Inserting $B$ into the action and collecting all the terms we get
\bea
\mathcal{L}&=& -\frac{1}{2\gamma g_{YM}^2}[W^{\alpha}_d + 2\gamma(
\Phi^i_{x}\nabla^{\alpha}\Phi^{i}_d + g_{YM}
\nabla^{\alpha}\Phi^{8}_d-\Phi^{i}_dD^{\alpha}\Phi^{ix})]^2 + 4\,k\,
\epsilon^{bcd}\,c_{jkl}\, g_{YM}\,\Phi^{j}_{b} \Phi^{k}_{c}
\Phi^{l}_{d} + \non\\
&&
+ \gamma \left[
\nabla^{\alpha}\Phi^i_d \nabla_{\alpha}\Phi^{id} +
\nabla^{\alpha}\Phi^8_d \nabla_{\alpha}\Phi^{8d} +
D^{\alpha}\Phi^i_{x} D_{\alpha}\Phi^{ix} \right] = \nonumber 
\\
&&
=\frac{1}{g_{YM}^2}W^{\alpha}_d W_{\alpha}^d
-\frac{1}{4}\nabla^{\alpha}\Phi^i_d \nabla_{\alpha}\Phi^{id} -\frac{1}{4}
D^{\alpha}\Phi^i_{x} D_{\alpha}\Phi^{ix} -\frac{1}{6}\,
\epsilon^{bcd}\,c_{jkl}\, g_{YM}\,\Phi^{j}_{b} \Phi^{k}_{c}
\Phi^{l}_{d} + ...\,,
\eea
where the dots stand for subleading contributions  and we substituded the values in (\ref{vallon}), so that now
\beq
W_{\beta}^d = \frac{1}{2} \, D^{\alpha}D_{\beta} A_{\alpha}^d +
\epsilon^{dba}(D^{\alpha}A_{\beta a})A_{\alpha\,b}\,.
\eeq
Rescaling the gauge superfield as in \cite{Papag1}  $A\rightarrow 1/2 A$
we see that $W^\alpha_d W_\alpha^d $ gives the right SYM kinetic term in the Wess-Zumino gauge and the covariant derivative in (\ref{covdev}) assumes its standard form.  Finally, ignoring the contribution of  $\Phi^i_x$ that has completely decoupled, we arrive at 
\beq
\label{SYM_final}
 \mathcal{L}= \frac{1}{8g_{YM}^2} (D^{\alpha}D_{\beta} A_{\alpha}^d +
\epsilon^{dba}(D^{\alpha}A_{\beta a})A_{\alpha\,b})^2
-\frac{1}{4}\nabla^{\alpha}\Phi^i_d \nabla_{\alpha}\Phi^{id}
-\frac{1}{6}\, \epsilon^{bcd}\,c_{jkl}\, g_{YM}\,\Phi^{j}_{b}
\Phi^{k}_{c} \Phi^{l}_{d}\,.
\eeq
 This is the superfield Langangian for maximally supersymmetric
YM in 2+1 dimensions for $SU(2)$.  It is intriguing to notice that the Lagrangian
(\ref{SYM_final}) is remarkably similar to the octonionic ${\cal
N}=(1,1)$ sigma model (in two dimensions) of \cite{Ferretti}.

\section{Conclusions}
We have presented an ${\cal N}=1$ superfield action in three
dimensions that in components gives the Bagger-Lambert action for a general 3-algebra with totally antisymmetric structure constants $f^{abcd}$. Crucial in our construction were the
self-dual octonionic tensors $C_{IJKL}$. Although the tensors are
$SO(7)$ invariant, we have shown that a special choice of the
parameters in the action enhances the global symmetry to $SO(8)$. We
have demonstrated that a superhiggs mechanism yields the maximally
supersymemtric 2+1 YM theory on D2 branes, curiously in a formalism
resembling a two-dimensional sigma model.  We hope that our superfield
action and its generalizations can be used in ${\cal N}=1$ superfield
calculations that should shed more light into the AdS$_4$/CFT$_3$
correspondence.

\subsection*{Acknowledgments}
We would like to thank A. Santambrogio for useful discussions. We are
also grateful to Hai Lin for very useful correspondence after the
first version of the paper. The work of A.C.P. is partially supported
by the European RTN Program MRTN-CT-2004-512194.

\begin{appendix}

\section{Superspace Notations}

In this appendix we collect the useful identities for our superfields and gamma matrices. We follow Superspace \cite{superspace}.

The component field definitions are as
$$
\begin{array}{rclrcl}
\Phi^I_a | & \equiv & \phi^I_a &\hspace{1cm} D_{\alpha}\Phi^I_a | &
\equiv & \psi^I_{\alpha \,a} \vspace{0,2cm} \non \\ D^2 \Phi^I_a | &
\equiv & F^I_a &\hspace{1cm} \Gamma_{\alpha \,ab} |& \equiv &
\chi_{\alpha\,ab}\vspace{0,2cm} \non \\ \frac{1}{2}\, D^{\alpha}
\Gamma_{\alpha\,ab} |& \equiv & B_{ab} &\hspace{1cm} D^2
\Gamma_{\alpha\,ab} | & \equiv & 2 \lambda_{\alpha\,ab} -
i\,\partial^{\,\beta}_{\alpha}\chi_{\beta\,ab}
 \vspace{0,2cm} \non \\D_{\alpha}\Gamma^{\beta}_{\,\,a b}| & \equiv &
i \left(\gamma_{\mu} \right)^{\,\beta}_{\alpha}
A^{\mu}_{\,\,a b} - \delta^{\,\beta}_{\alpha} B_{a b} & \hspace{1cm} D^2
\Gamma^{\alpha}_{\,\,a b} | & \equiv & 2 \lambda_{\,\,a b}^{\alpha} +
i\,\partial_{\,\,\beta}^{\alpha}\chi_{\,\,a b}^{\beta} \vspace{0,2cm} \\
\hspace{1cm} D^{\alpha}\Gamma_{\beta \,a b}| & \equiv &
i \left(\gamma_{\mu} \right)_{\,\beta}^{\alpha}
A^{\mu}_{\,\,a b} + \delta_{\,\beta}^{\alpha} B_{a b} &\hspace{1cm}
\frac{1}{2} D^{\beta}D_{\alpha} \Gamma_{\beta \, ab}|& \equiv &
\lambda_{\alpha \, ab} \non
\end{array}
$$
In the analog of the Wess-Zumino procedure we can gauge away $B$ and $\chi$
component fields.
Our spacetime signature is $(-,+,+)$. The purely immaginary
totally antisymmetric symbol $C_{\alpha\beta}$ is used to raise and lower
spinor indexes according to the $\searrow\,$ convention
\begin{eqnarray}
C_{\alpha \beta} & = & - C_{\beta \alpha} = -
C^{\alpha \beta} = \left(\begin{array}{cc} 0 & -i \\ i & 0
\end{array}\right) \qquad \!\!\!\!\qquad C_{\alpha \beta}C^{\gamma
\delta} = \delta^{\gamma}_{[\alpha} \delta^{\,\delta}_{\beta]} \non \\
\psi_{\alpha} &=& \psi^{\beta}C_{\beta \alpha} \qquad
\,\,\,\,\,\,\qquad\psi^{\alpha} = C^{\alpha \beta} \psi_{\beta} \qquad
\,\,\,\,\,\,\qquad \psi^2 = \frac{1}{2} \psi^{\alpha}\psi_{\alpha}
 \non
\end{eqnarray}
We represent vectors in spinor notation as
symmetric matrices:
$$
\begin{array}{ccccccccc}
(\gamma^0)_{\alpha\beta}
& = & - (\gamma^0)^{\alpha\beta}& = & - (\gamma_0)_{\alpha\beta}& = &
(\gamma_0)^{\alpha\beta}  & = & \left(\begin{array}{cc} 1 & 0 \\ 0 & 1
 \end{array}\right) 
 \vspace{0,2cm}\non \\ (\gamma^1)_{\alpha\beta}
& = & (\gamma^1)^{\alpha\beta}& = & (\gamma_1)_{\alpha\beta}& = &
(\gamma_1)^{\alpha\beta} & = &\left(\begin{array}{cc} 1 & 0 \\ 0 & -1
 \end{array}\right) 
 \vspace{0,2cm}\non \\(\gamma^2)_{\alpha\beta}
& = & (\gamma^2)^{\alpha\beta}& =&  (\gamma_2)_{\alpha\beta}&=&
(\gamma_2)^{\alpha\beta}  &=& \left(\begin{array}{cc} 0 & 1 \\ 1 & 0
 \end{array}\right) 
 \vspace{0,2cm}\non \end{array} 
$$
$$
(\gamma^{\mu})_{\alpha\beta}V_{\mu} =
\left(\begin{array}{cc} V_0+ V_1 & V_2 \\ V_2 & V_0-V_1
 \end{array}\right)  
$$
Then the following relations hold:
$$
\begin{array}{rclrcl}
 \partial^{\alpha
\beta} & = & \left(\gamma_{\mu}\right)^{\alpha \beta} \partial^{\mu}
&\hspace{.5cm}\partial^{\mu} & = &\frac{1}{2} \,
\left(\gamma^{\mu}\right)_{\alpha \beta} \partial^{\alpha
\beta}\vspace{0,2cm}\non \\ x^{\alpha \beta} & = &
\frac{1}{2}\left(\gamma_{\mu}\right)^{\alpha \beta} x^{\mu}
&\hspace{.5cm} x^{\mu} & = &\left(\gamma^{\mu}\right)_{\alpha \beta}
x^{\alpha \beta}
\vspace{0,2cm}\non \\ A^{\alpha \beta} & = &
\frac{1}{\sqrt{2}}\left(\gamma_{\mu}\right)^{\alpha \beta} A^{\mu}
&\hspace{.5cm}A^{\mu} & =
&\frac{1}{\sqrt{2}}\left(\gamma^{\mu}\right)_{\alpha \beta}
A^{\alpha \beta} \non \vspace{0,2cm} \\ \partial_{\alpha \beta}
x^{\gamma \rho}& = & \frac{1}{2} \,
\delta^{\,\,\gamma}_{(\alpha}\delta^{\,\,\rho}_{\beta)} &
\hspace{.5cm}\partial_{\alpha}\theta^{\beta} & =
&\delta_{\alpha}^{\,\,\beta}\non \\(\gamma_{\mu})^{\alpha
\beta}(\gamma_{\nu})_{\alpha \beta} & = & 2 \eta_{\mu \nu} &
\hspace{.5cm} (\gamma^{\mu})_{\alpha \beta}(\gamma_{\mu})^{\gamma
\rho} & = & \delta_{\alpha}^{\,\,\gamma} \delta_{\beta}^{\,\,\rho} +
\delta_{\alpha}^{\,\,\rho} \delta_{\beta}^{\,\,\gamma} \vspace{0,2cm}
\non\\ (\gamma^{\mu})^{\alpha \beta}(\gamma^{\nu})_{\beta \rho} & = &
\eta^{\mu \nu}\delta^{\alpha}_{\rho} + i
\epsilon^{\mu\nu\sigma}(\gamma_{\sigma})^{\alpha}_{\,\,\,\rho} &
(\gamma^\mu)^{\rho}_{\,\,\rho}& = & 0 \vspace{0,2cm} \non\\(\gamma^{\mu})^{\alpha
\beta}(\gamma^{\nu})_{\beta}^{\,\,\rho}(\gamma^{\sigma})_{\rho\alpha}
& = & 2i \epsilon^{\mu\nu\sigma} &\hspace{.5cm} \epsilon^{012} & = & 1
\vspace{0,2cm} \non

\end{array}
$$
\noindent
We conclude with some useful relations for three dimensional D-algebra
computations: \\
$$
\begin{array}{rclrcl}
D^{\alpha}D_{\beta} & = & i
\,\partial^{\alpha}_{\,\,\beta} +
\delta^{\alpha}_{\,\beta} \,D^2 & \hspace{3cm}
D_{\alpha}D^{\beta} & = & i\,
\partial_{\alpha}^{\,\,\beta} -
\delta_{\alpha}^{\,\beta}\, D^2
\vspace{0,2cm}\non \\ D^2 D_{\alpha} & = & - i
\,\partial_{\alpha}^{\,\,\beta} \, D_{\beta} &\hspace{3cm} D_{\alpha}D^2 &
= & i \,\partial_{\alpha}^{\,\,\beta} \, D_{\beta} \vspace{0,2cm} \non \\
D^2 D^{\alpha} & = & i \,\partial^{\alpha}_{\,\,\beta} \, D^{\beta}
&\hspace{3cm} D^{\alpha}D^2 & = & - i \,\partial^{\alpha}_{\,\,\beta} \,
D^{\beta} \vspace{0,2cm} \non \\ D^{\alpha} D_{\beta} D_{\alpha} & = & 0 &
\hspace{3cm} \partial^{\alpha \beta} \partial_{\gamma \beta} & = &
\delta^{\alpha}_{\,\gamma} \, \Box \vspace{0,2cm} \non \\ D^2 D^2 & = &
\Box & \hspace{3cm} D^{\alpha} D^2 D_{\alpha} & = & -2 \, \Box
\vspace{0,2cm} \non \\ D^2 D_{\alpha} D^2 & = & - \, \Box \,
D_{\alpha} & \hspace{3cm} \Box & = & \partial^{\mu}\partial_{\mu} \, =
\,\frac{1}{2}\partial^{\alpha \beta} \partial_{\alpha \beta}
\vspace{0,2cm} \non
\end{array}
$$\\
\newpage
\noindent

\section{Octonionic conventions}
In the octonion algebra we can choose a basis of elements
 $$\Big\{1,\, e_i\Big\} \qquad \qquad i=1,..,7$$
such that 
$$
e_i e_j=c_{ijk}\,e_k - \delta_{ij}
$$
where the tensor $c_{ijk}$ is totally antisymmetric with non vanishing entries:
$$c_{123}=c_{147}=c_{165}=c_{246}=c_{257}=c_{354}=c_{367}= 1$$ 
We can
also introduce the seven dimensional dual of the structure constants:
$$
c_{ijkl} = \frac{1}{6} \,\epsilon_{ijklmno}\,c^{mno}
$$ Combining these two objects one can construct an $SO(7)$ invariant
tensor $C_{IJKL}\,\,_{I,J,K,L=1,...,8}$ which is self dual in 8
dimensions, by taking:
$$
C_{ijk8}=c_{ijk} \qquad \qquad C_{ijkl}=c_{ijkl}
$$ Octonionic structure constants can be used to construct $SO(8)$
 gamma matrices. For instance a suitable
 representation of the triality tensor that enters BL susy
 transformations is given by:
\begin{eqnarray}
 (\Gamma^i)_{A\dot{A}} &=& c^i_{\,\,A\dot{A}} +
\delta_{8\dot{A}}\delta_{Ai} - \delta_{8A} \delta_{\dot{A}i} \qquad
\qquad i=1,...,7 \qquad A,\dot{A}=1,...,8 \non\\ (\Gamma^8)_{A\dot{A}}
& = & \delta_{A\dot{A}} \qquad \qquad \qquad \qquad\qquad \qquad
c^{i}_{8\dot{A}} = c^{i}_{A8}= 0\non
\end{eqnarray}
Defining $\hat{\Gamma}^I_{\dot{A}A}=\left(\Gamma^{T}\right)^I_{\dot{A}A}$ it is
easy to see that $\Gamma^I\hat{\Gamma}^J + \Gamma^J\hat{\Gamma}^I =
2\delta^{IJ}$ and therefore we can write down 16x16 matrices
satisfying the clifford algebra
$$
\gamma^I=\left(\begin{array}{cc}0 & \Gamma^I_{\,\,A\dot{A}} \\
\hat{\Gamma}^I_{\,\,\dot{A}A} & 0\end{array}\right)\,,\hspace{2cm}
\gamma^I\gamma^J+\gamma^J\gamma^I= 2 \delta^{IJ}\,.
$$
With the above definitions  it can be shown  that
$$
\Gamma^{IJ}_{\,\,AB} =\frac{1}{2}\left(\Gamma^I_{A\dot{A}}\hat{\Gamma}^J_{\dot{A}B}-\Gamma^J_{A\dot{A}}\hat{\Gamma}^I_{\dot{A}B}\right)= C^{IJ}_{\,\,\,AB} +\delta^I_A \delta^J_B
-\delta^I_B \delta^J_A\,.
$$
 so that the antisymmetrized product of $\Gamma$'s of the
scalar-fermion interaction can be written by means of the $C_{IJKL}$
tensor.
\end{appendix}

\end{document}